\documentclass[prl,twocolumn,floatfix,showpacs]{revtex4}%
\usepackage{bm}		
\usepackage{amsmath}
\usepackage{amssymb}
\usepackage{graphicx} 
\newcommand{\EPSwill}[3]{        
\begin{figure}
\includegraphics[width=#1cm]{#3} 
     	\caption{\label{#3} #2}
     \end{figure}
}
\newcommand{\bn}{\bm n}
\newcommand{\br}{\bm r}
\newcommand{\bu}{\bm u}
\newcommand{\bff}{\bm f}
\newcommand{\bdot}{\bm\cdot}
\newcommand{\av}[1]{\langle#1\rangle}
\newcommand{\sd}{\mbox{d}}

\newcommand{\fig} [1]{Fig.~\ref{#1}}
\newcommand{\Fig} [1]{Figure~\ref{#1}}
\newcommand{\figs}[2]{Figs.~\ref{#1} and~\ref{#2}} 

\begin{document}
\title{Dynamic scaling in stick-slip friction}
\author{Jens \surname{Feder}}
 \email{feder@fys.uio.no}
\author{H{\aa}kon \surname{Nordhagen}}
\author{Wesley Andr{\'e}s \surname{Watters}}
\altaffiliation[Also at: ]{Department of Earth Atmospheric
     and Planetary Sciences, MIT, 
     Cambridge, MA 02139, U.S.A.}
\affiliation{Physics of Geological Processes, 
University of Oslo, Box 1048 Blindern,
N-0316 Oslo, Norway}
\date{\today}

\begin{abstract}
We introduce a generalized homogeneous function to describe the joint
probability density for magnitude and duration of events in
self-organized critical systems (SOC). It follows that the cumulative
distributions of magnitude and of duration are power-laws with
exponents $\alpha$ and $\tau$ respectively. A power-law relates
duration and magnitude (exponent $\gamma$) on the average. The
exponents satisfy the dynamic scaling relation $\alpha=\gamma\tau$.
The exponents classify SOC systems into universality classes that do
not depend on microscopic details provided that both $\alpha\le1$ and
$\tau\le1$.
We also present new experimental results on the stick-slip motion of a
sandpaper slowly pulled across a carpet that are consistent with our
criteria for SOC systems. Our experiments, as well as experiments by
others, satisfy our dynamic scaling relation. We discuss the relevance
of our results to earthquake statistics.
\end{abstract}
\maketitle

\section{Introduction}
The force required to pull a block along a surface is proportional to
its weight and independent of contact area.  The sliding motion of the
block depends on the elastic properties of the string pulling the
block, the block itself, and the surface, and takes many forms
\cite{BT54-68,Per98,GM01}.  For a stiff system, periodic slips with
a velocity dependent amplitude have been observed
\cite{BHP94,HBP*94}. Intermittent stick-slip friction has been
observed in experiments where a highly compliant string was used to
pull a piece of sandpaper across a soft carpet \cite{FF91}. We extend
these measurements to resolve the slip duration and find that our
system is an example of a self-organized critical (SOC) system
\cite{BTW87}.

We propose a testable criterion for scale-free SOC systems: The joint
 probability density of slip magnitude and duration is a generalized
 homogeneous function, the exponents of the cumulative magnitude and
 duration distributions must be less than unity, so that moments of
 the distributions do not exist, and a dynamic scaling relation
 between the exponents must be satisfied. It follows that periodic
 stick-slip motion \cite{BHP94,HBP*94} cannot be SOC. The experimental
 results presented here satisfy the definition of SOC accurately.

The scaling relation is also satisfied by the magnitude and duration
exponents of rain showers \cite{PC02}, and by exponents for slips in a
granular medium \cite{DC01}. In our view, earthquakes are not SOC
since the currently accepted exponents are larger than one.

\section{Experiments}
A weighted sandpaper was pulled by an elastic nylon line slowly (9
$\mu$m/s) across a carpet \cite{methods}.  The sandpaper moved by a
stick-slip process.  We measured the force required to hold the carpet
in place (see \fig{foft}) at intervals of $\delta t = (14
\text{Hz})^{-1}$, that is, every 0.6 $\mu$m on the average. For each
slip, we observed an abrupt decrease in force, $\Delta F$, and for the
first time we have been able to measure the duration, $t$, of slips.

The force per unit area (stress), $\bff(\br)$, between the sandpaper
and the carpet is a function of position, $\br$, within the surface
area, $A$, of the sandpaper. The integral of the vertical component of
$\bff(\br)$ is a constant equal to the weight of the sandpaper 
slider (mass $m$): $mg =\int_A\bff(\br)\bdot\bn\, \sd^2\br$, where $g$ 
is the acceleration of gravity, and $\bn$ is the surface normal.  In a slip,
the contact between the sandpaper and the carpet is lost, or reduced,
over an area $S\leq A$, that is not necessarily connected. The force
changes by $\delta\bff(\br)= \mu\;\bu(\br)/h $, where $\bu(\br)$ is
the local slip vector, in the plane of the carpet, $h$ is the carpet
thickness, $\mu$ is the shear modulus of the carpet, and
${\bu(\br)/h}$ is the slip-associated change in shear strain at 
$\br$. The slip leads to an elastic increase in shear strain, $\bu_e(\br)/h$
outside $S$, and increased elastic stress takes more of the external
load. Integrating $ \delta\bff(\br)$ over the area of the sandpaper,
$A$, we find that
\begin{equation*}
  \begin{split}
    m= h\Delta F&= h \int_A\delta\bff(\br)\bdot\sd^2\br\\
    &= \mu\int_S \bu(\br)\bdot\sd^2 \br 
   + \mu\int_{A-S} \bu_e(\br)\bdot\sd^2 \br \\
&= \mu' S \av{u}_S\;.
\end{split}
\end{equation*}

  Experimentally we measure the first integral. The next integrals
  represent a model where slip occurs over an area $S$, and an
  increased elastic strain elsewhere in $A$.  The integral over $A-S$
  is proportional to the integral over $S$, and $\av{u}_S$ is the
  mean slip averaged over the slip area $S$. The last expression is
  the standard form of the (scalar) seismic moment $M_0$ with an
  effective shear modulus $\mu'$ \cite{Sch02,KB04}. We conclude that
  by measuring $\Delta F$ we obtain the seismic moment, $m$, of the
  slip directly.

\EPSwill{8}{Force as a function of time from an experiment.  The
  initial rise and drop in force at the beginning of the experiment is
  characteristic of the interval of time in which the system
  ``forgets'' its initial state.  Later slips of size $\Delta F$ and
  duration $t$, are studied here.}{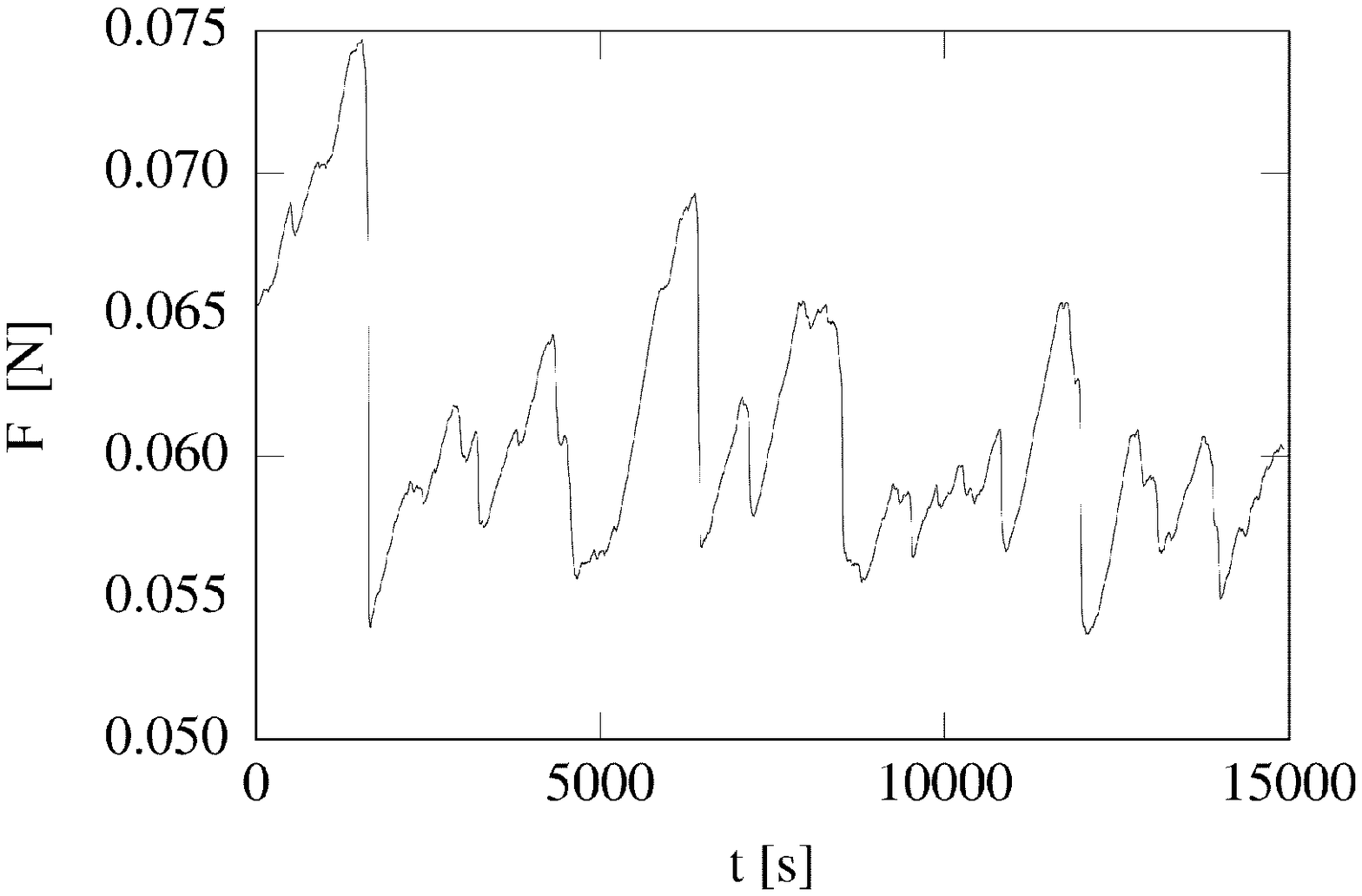}

\EPSwill{8}{The cumulative distribution, $P(M > m)$, as a function of
the seismic moment, $m$.  The observed cumulative distribution is
consistent with a power-law, $P(M > m)\sim m^{-\alpha}$, with
$\alpha=0.67\pm0.03$. The straight line has a slope of $-0.67$.}{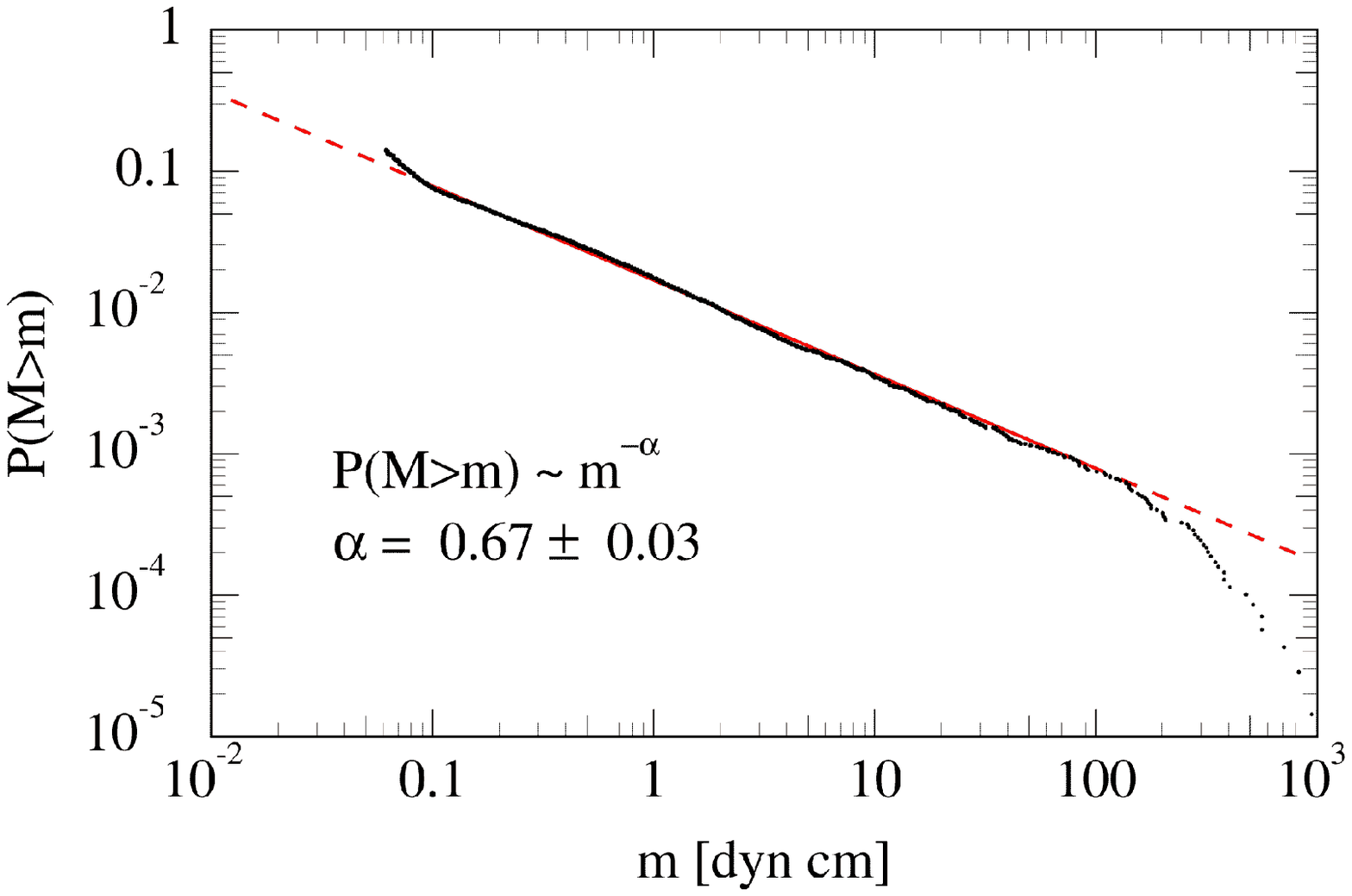}

A slip starts at $t_1$ if the force difference between two consecutive
measurements $\delta F(t_1) = F(t_1+t)-F(t_1)$, changes sign from
positive to negative at $t_1$, and ends at $t_2=t_1 +t$ when $\delta
F(t_2)$ becomes positive. A slip is therefore characterized by the
decrease in force $\Delta F=F(t_1)-F(t_2)$, that is, by the seismic
moment $m=h\Delta F$ and duration $t=t_2-t_1$.  We found that this
procedure gave the same scaling exponents as more complicated ways of
estimating slip magnitude and duration, such as sliding average and
other filtering procedures, that we have tested. Of course, each slip
consists of events whereby fibers lose their grip on the sandpaper
one by one; the sound from slips was easy to hear. From this point of
view the minimum slip consist of one single fiber losing its grip.  
We have seen that there are bursts of such micro-slips in the slips identified
by us. There was some very low activity even in the intervals between
slips. Of course, electrical noise from the strain-gauge, and
measurement system limits the resolution so that we cannot separate
the smallest slips from electrical noise. For the carpet used in
the experiments reported here, in no case did all fibers lose their 
grip on the sandpaper, so that the sandpaper was never displaced more
than one fiber length with respect to the carpet. In the case of 
carpets for which this did sometimes occur, a power-law distribution 
of magnitudes and durations was not observed.

\Fig{CumMX} shows that the cumulative distribution of magnitude,
averaged over four independent experiments with a total of more than
50\,000 slips, is consistent with the scaling form \cite{SlipDef}
\begin{equation}\label{eqcdm}
P(M > m) \sim m^{-\alpha}\; ,\qquad \alpha=0.67\pm0.03\;.
\end{equation}
%

The deviations from the power-law fit (dashed line segments were not
part of the fit) at large $m$ are due to finite-size effects, and the
increase at small $m$ is due to noise, which masquerades as small
slips.  Similar deviations for small and large values are seen in
\figs{CumT}{TvsM}. Since power-law distributions are stable under
addition \cite{FGN}, the effect of noise and finite sampling rate does
not affect the slope of the distributions, but reduces the range over
which scaling may be observed.
\EPSwill{8}{ The cumulative distribution, $P(T > t)$, as a function of
slip duration $t$. The observed cumulative distribution is consistent
with a power-law, $P(T >t)\sim t^{-\tau}$, with $\tau=0.96\pm
0.03$. The straight line has slope $-0.96$.}{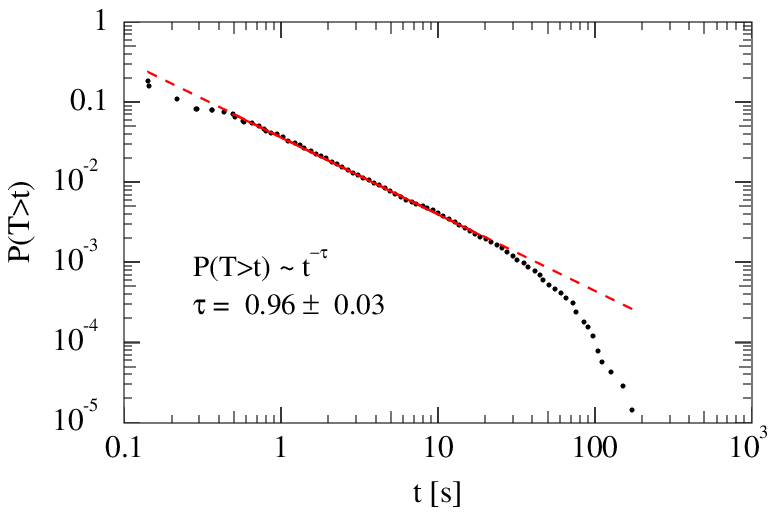}

\EPSwill{8}{The joint probability density, $p(m,t)$, of seismic
moment, $m$, and slip duration, $t$, is illustrated by showing the
most probable values $(m_0,t_0)$ as points. The shaded area represents
$(m,t)$ values within one standard deviation around $(m_0, t_0)$. The
straight line is a fit of $t_0 \sim m_0^\gamma$ with a slope
$\gamma=0.69$.}{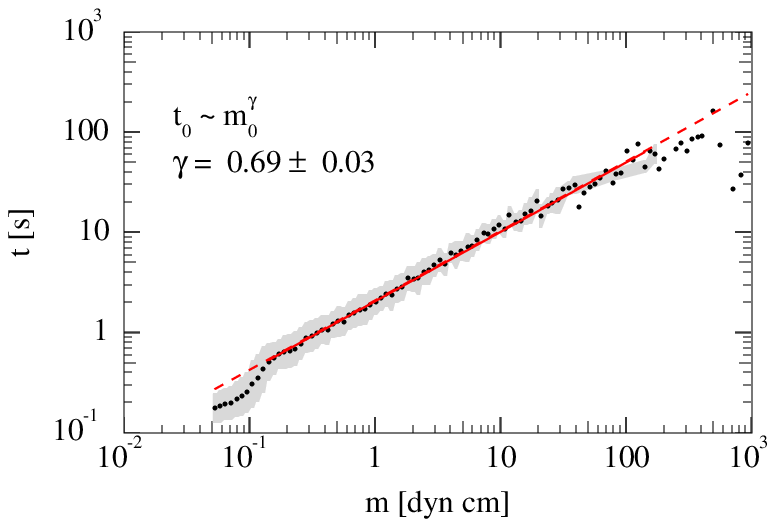}

\Fig{CumT} shows that the cumulative distribution of event duration
 time $t$ is consistent with the scaling form
\begin{equation}\label{eqcdt}
P(T>t) \sim t^{-\tau}\;,\qquad\tau=0.96\pm0.03\;.
\end{equation}

We also found that slips with a large seismic moment lasted longer
than small slips.  \Fig{TvsM} shows the estimates of most probable
values $(m_0,t_0)$ of the joint probability density $p(m,t)$ as
points, and estimates of the standard deviations (for bins that
contain more than six measurements) around these points as the shaded
area, that is, we have estimated $t_0$ to be the average duration given
$m=m_0$.

The results shown in \fig{TvsM} are consistent with the power-law form
relating the most probable pairs $(m_0,t_0)$:
\begin{equation}\label{eqjoint}
t_0 \sim m_0^{\gamma}\;,\qquad \gamma=0.69\pm0.03\;,
\end{equation}
for $m > 0.15$ dyn\,cm.


\section{Theory}
SOC, as originally defined \cite{BTW87}, requires power-law scaling
both in magnitude and duration of the events.  We show that assuming
the joint probability density $p(m,t)$ to be a generalized homogeneous
function, it follows that the magnitude and duration distributions are
power-laws, and average duration as a function of magnitude is also a
power-law.  The measured exponents satisfy a scaling relation that
follows from the assumption that the joint probability density,
$p(m,t)$, is a generalized homogeneous function, that is, for any
positive value of $\lambda$:
\begin{equation}
p(m,t)\sd m\, \sd t = \lambda^{-\alpha}p\left(
  \frac{m}{\lambda},\frac{t}{\lambda^{\gamma}}\right)
\sd\left(\frac{m}{\lambda}\right)\sd\left(\frac{t}{\lambda^{\gamma}}\right)\;.
\label{eqhomog}
\end{equation}
Here $\lambda$ is a number that represents the change of scale
provided $m$ and $t$ are given in dimensionless units. Such functions
have been discussed extensively in the context of scaling at the
critical point of second order phase transitions
\cite{Domb96}. However, for non-equilibrium systems, of the kind
discussed here, the theoretical basis for (\ref{eqhomog}) has not been
established and it thus remains a proposal.

The probability density, $p(m)$, for slip magnitude is obtained from
eq. (\ref{eqhomog}) by integrating over the slip duration
\[
\begin{split}
p(m)\sd m &=\int_{t=0}^\infty p(m,t)\sd m\,\sd t \\
          &=\lambda^{-(1+\alpha)}\sd m\int_{v =0}^\infty
               p\left( \frac{m}{\lambda},v\right) \sd v \;,\\
\end{split}
\]
where we have introduced the integration variable
$v=t/\lambda^\gamma$. Since $\lambda$ may be chosen to have any
positive value we set $\lambda=m$, and find
$$
p(m)=m^{-(1+\alpha)}\int_{v=0}^\infty p(1,v)\sd v \sim m^{-(1+\alpha)}\;.
$$
 Here the integral is a constant, and it follows that the cumulative
distribution, $P(M > m)=\int_m^{\infty}p(M)\,\sd M\sim m^{-\alpha}$,
has the scaling form given in eq. (\ref{eqcdm}).

The probability density, $p(t)$, for the slip duration $t$ is found, in
a similar way, by integrating eq. (\ref{eqhomog}) over $m$:
$$
p(t)=t^{-(1+\alpha/\gamma)}
\int_{v=0}^\infty p(v,1)\,\sd v \sim t^{-(1+\alpha/\gamma)}\;.
$$ 
Again the integral is a constant, and it follows that the cumulative
distribution, $P(T > t)=\int_{t}^{\infty}p(T)\, \sd T\sim
t^{-\alpha/\gamma}$, has the scaling form given in eq. (\ref{eqcdt})
with $\tau=\alpha/\gamma$, that is, we have the dynamic scaling relation
\begin{equation}\label{eqscaling}
\alpha=\gamma\tau\;.
\end{equation}
This relation also follows by assuming the scaling forms
(\ref{eqcdm})--(\ref{eqjoint}) and requiring consistency for the
conditional expectation values \cite{CFJ91}.

Equation (\ref{eqjoint}) follows from our proposition
(\ref{eqhomog}).  Let $(m_0^*,t_0^*)$ be a $(m, t)$ pair where
$p(m,t)$ has a local maximum corresponding to the most\maketitle probable value
for $m$ given $t=t_0^*$ or $t$ given $m=m_0^*$. Then, for the
generalized homogeneous function (\ref{eqhomog}), $m_0=m_0^*/\lambda$
and $t_0=t_0^*/\lambda^\gamma$ also corresponds to a local maximum in
the joint probability density for any positive value of $\lambda$; it
follows that
$$
1/\lambda^{\gamma}=t_0/t_0^* 
= \left({m_0}/{m_0^*}\right)^\gamma \;,
$$
which is indeed eq. (\ref{eqjoint}).  Since both $m$ and $t$ are taken
to be dimensionless, we may chose both $m_0^*$ and $t_0^*$ to be unity
for the purposes of the scaling relations. Equation (\ref{eqhomog}),
states that the joint probability density is invariant with respect to
the affine transformation $m\to m/\lambda$, $t\to t/\lambda^\gamma$,
and $p\to p/\lambda^\alpha$; it follows that any expectation and other
characteristic, such as the full width at half height, will scale and
have exponents that only depend on $\alpha$ and $\gamma$. It follows
that the width of the distribution in \fig{TvsM} should be constant in
the log-log plot.

If there is a characteristic magnitude, or duration, of slip events,
then any model of the process must include the characteristic values.
However, if $\alpha \le 1$ and $\tau \le 1$,  no characteristic
value exists.  Consider the expectation value of the magnitude
\begin{equation}
\begin{split}
\av{m}&=\int_{m_\epsilon}^{m_L}m\,p(m)\sd m
\sim\int_{m_\epsilon}^{m_L}m^{-\alpha}\sd m\\
&=\frac{m_L^{1-\alpha}-m_\epsilon^{1-\alpha}}{1-\alpha}
\to\infty
\begin{cases}
 \alpha<1 \text{ and } m_L\to\infty\;,\\
\alpha>1 \text{ and } m_\epsilon\to 0\;.\\                
\end{cases}\\
\end{split}
\label{eqmoment}
\end{equation}
Here $m_L$ is the maximum event magnitude, which diverges with system
size $L$, and $m_\epsilon$ is the smallest, ``atomic scale", event
size.  If $\alpha\le 1$, then $\av{m}$ fails to exist and diverges
with system size (note that $\av{m}\to \ln{m_L}$ as $\alpha\to 1^-$).
Similarly, if $\tau\le 1$ then $\av{t}\to\infty$ as system size
increases and the system is also scale-free in the time domain.  In
this case the system is {\em scale-free} and has no characteristic
event magnitude or duration; and any model of the system must also be
scale-free and cannot depend on atomistic details. We consider such
models not to be SOC.

We therefore consider systems to be self-organized critical provided
that eqs. (\ref{eqhomog}) and (\ref{eqscaling}) are satisfied, and the
first and higher moments of both the magnitude- and the duration
probability densities diverge with system size.

From the modeling point of view, we expect that systems with finite
first moments, that is, $\alpha>1$, and/or $\tau>1$, require
``atomistic'' details since the distributions then are dominated by
the small scale and/or short time behavior, i.e., dominated by
$m_\epsilon$ in (\ref{eqmoment}) even if $m_L \to \infty$.

\begin{table}[!h]
\caption{\label{table1} Scaling exponents for magnitude and
duration\hfill}
\label{t.1}
\begin{center}
\begin{tabular}{lccccc}\hline
Distribution && Carpet & Rain & Granular& Earthquakes\\
 &&  & \cite{PC02}& \cite{DC01}& \cite{Sch02,KB04}\\\hline
$P(M>m)\sim m^{-\alpha}$ & $\alpha$& 0.67   & 0.4  &   0.94   & 2/3  \rule{0pt}{5mm}\\
$ P(T>t)\sim t^{-\tau}$ & $\tau$   & 0.96   & 0.6  &   1.08   & 2     \\
$t\sim m^{\gamma}$     & $\gamma$  & 0.69   & 0.7  &   0.87   & 1/3 \\\hline
$\alpha = \gamma\tau$& $\gamma\tau$& 0.66   & 0.4  &   0.94   & 2/3 \rule{0pt}{5mm}\\\hline
\end{tabular}
\end{center}
\end{table}
 
\section{Discussion}
Table \ref{table1} collects values for the scaling exponents and tests
the validity of the dynamic scaling relation eq. (\ref{eqscaling}). In
our experiments we have measured the exponents $\alpha$, $\tau$, and
$\gamma$. We find that both $\alpha<1$ and $\tau<1$, and that the
scaling relation $\alpha=\gamma\tau$ holds within experimental
accuracy. We conclude that the stick-slip process studied is SOC.

The dynamic scaling relation may also be tested on other ``avalanche
phenomena'' where size and duration are known to be power-law
distributed such as the size and duration of rain showers \cite{PC02},
and the slips in a granular medium under shear \cite{DC01}. We note
that the granular slip experiments \cite{DC01} have $\tau=1.08$ and
the system is therefore not scale free unless $\tau \le 1$, which is
within experimental error.

The exponents take different values for different systems. In the
  language of critical phenomena \cite{Domb96}, we would say that
  stick-slip motion and rain showers belong to different
  universality classes, since the exponents differ. We expect that for
  each SOC universality class there exist a scale free dynamic model
  characterizing that class.

Earthquakes have long been recognized as resulting from stick-slip
  friction \cite{Sch98}. Earthquakes may represent an example of SOC
  \cite{BTW87,BT89,BCDS02}.  The magnitude of earthquakes is
  characterized by their {\em seismic moments} $M_0$ \cite{Sch02},
  which has the Gutenberg-Richter distribution \mbox{$P(M>M_0)\sim
  M_0^{-B}$} with the exponent $B=\alpha = 2/3$
  \cite{Sch02,KB04}. Implicitly it is assumed that $M_0$ is the only
  stochastic variable for earthquakes, other quantities such as
  duration $t$, and the area $S$ are ``mechanically'' related to $M_0$.
  The duration is $t=L/V$, where $V$ is the rupture velocity, and $L
  =S^{1/2}$; also $M_0\propto S^{3/2}$. These scaling relations are
  consistent with observations \cite{KB04}. It follows that $t\propto
  M_0^{1/3}$, i.e., $\gamma=1/3$; and $\tau=2$. Consequently, in our
  view, earthquakes are not SOC since they are not described by a
  joint probability density for magnitude and duration that is a
  generalized homogeneous function, and because $\tau>1$. Models of
  earthquakes must include ``microscopic'' details for the duration,
  and be scale free for magnitudes, in contradiction with the
  mechanistic relation $t= M_0^{1/3}$.

The original paper \cite{BTW87} on SOC introduced a cellular automaton
model for avalanches (with $\alpha=0$ for a $(50)^2$ system). However,
the exponents are not well determined since they depend strongly on
system size \cite{CO93}. For a $(2048)^2$ system $\alpha=0.13$,
$\gamma=0.61$, and $\tau=0.20$ consistent with eq.~(\ref{eqscaling}).
The conservative OFC model \cite{OFC92} has exponents $\alpha=0.25$,
$\gamma=0.42$, and $\tau=0.52$ satisfying the scaling relation. Here
duration was taken to be the number of cellular automaton updates
required to end an avalanche.  Both models are SOC with our
definition. The models just discussed cannot be in the universality
class of our stick-slip motion, or earthquakes, since they have too
small values for $\alpha$. 

\section{Conclusion}
We conclude that stick-slip motion in the system we have studied
 exhibits power-law scaling that is well described by a joint
 probability density for magnitude and duration that has the form of a
 generalized homogeneous function.
We derived a scaling relation between the dynamic exponents $\tau$ and
  $\gamma$, and the magnitude exponent $\alpha$ that takes the simple
  form $\alpha=\gamma\tau$. We have found other examples in natural
  and experimental systems that satisfy this scaling relation. We show
  that our analysis leads to open questions for earthquakes.

We have proposed that self-organized critical (SOC) systems are
  characterized by
\begin{itemize}
\item a joint probability density for magnitude and duration that is a
  generalized homogeneous function,
\item with exponents $\alpha \le 1$ and $\tau \le 1$, and 
\item that satisfy the scaling relation $\alpha=\gamma\tau$.
\end{itemize}

\acknowledgments We thank Kim Christensen and Y. Kagan for
discussions.  WAW thanks the Fulbright foundation for a fellowship.


\bibliography{carpet}

\begin{thebibliography}{21}
\bibitem[{Sli()}]{SlipDef}
\bibinfo{note}{In reference \cite{FF91} the average speed was $270\, \mu$m/s,
  compared with only $9\,\mu$m/s here. The magnitude exponent was
  $\alpha=0.79\pm0.05$, which is above, but consistent with, the value found in
  the extensive experiments presented here.}
\end{thebibliography}


\end{document}